\documentclass{ws-ijmpc}

\usepackage{ulem}
\usepackage{hyperref}
\usepackage{lipsum,verbatim}
\usepackage{graphicx}
\usepackage{multirow,rotate,subfigure}
\usepackage{amsmath,amssymb}
\usepackage{booktabs}% for \toprule, \midrule, and \bottomrule

\usepackage[left]{lineno} 
\begin{document}

%\linenumbers

%\markboth{}
\markboth{Owen J.~Brison and Jason A.C.~Gallas}
{Polynomial interpolation as detector of orbital equivalence}

%%%%%%%%%%%%%%%%%%%%% Publisher's Area please ignore %%%%%%%%%%%%%%%
\catchline{}{}{}{}{}
%%%%%%%%%%%%%%%%%%%%%%%%%%%%%%%%%%%%%%%%%%%%%%%%%%%%%%%%%%%%%%%%%%%%

\title{Polynomial interpolation as detector of orbital equation equivalence}

%\author{}
\author{Owen J.~Brison$^1$ \ \  and \ \ Jason A.C.~Gallas$^{2,3,4}$}
\address{$^1$Departamento de Matem\'atica, %Faculdade de Ci\^encias,
  Universidade de Lisboa, 1649-003 Lisboa, Portugal,\\
%\author{Jason A.C.~Gallas} %\footnote{
% Typeset names in 8~pt Roman, upper and lower case.
% Use the footnote to indicate the
% present or permanent address of the author.}}
% \address{}
% \begin{comment}
$^2$Instituto de Altos Estudos da Para\'\i ba,
  Rua Silvino Lopes 419-2502,\\     %\\%\\ %58039-190 Jo\~ao Pessoa, Brazil
  58039-190 Jo\~ao Pessoa, Brazil,\\
$^3$Complexity Sciences Center, 9225 Collins Ave.~1208, 
        Surfside FL 33154, USA,\\
$^4$Max-Planck-Institut f\"ur Physik komplexer Systeme,
  % Max-Planck Institute for the Physics of Complex Systems,
  % N\"othnitzer Str.~38,
             01187 Dresden, Germany}%\\
%jason.gallas@gmail.com}
% \end{comment}

\maketitle

\begin{history}
\received{8 August 2018}
%\revised{Day Month Year}
\accepted{24 August 2018}
\centerline{Published 19 September 2018}
\centerline{https://doi.org/10.1142/S0129183118500961}
\end{history}

\begin{abstract}
  Equivalence between algebraic equations of motion may be detected by
  using a $p$-adic method, methods using factorization and linear algebra,
  or by systematic computer search of suitable Tschirnhausen transformations.
  Here, we show standard {\sl polynomial interpolation} to be a competitive
  alternative  method for  detecting orbital equivalences and field
  isomorphisms.
  Efficient algorithms for ascertaining equivalences are relevant for
  significantly
  minimizing computer searches in theoretical and practical applications.\\ 
\phantom{xxx}
%\hfill Draft of \today{} at \writetime

\end{abstract}

\ccode{PACS Nos.:
      02.70.Wz, %Symbolic computation (computer algebra)
      02.10.De, %Algebraic structures and number theory     
      03.65.Fd} %Algebraic methods (see also 02.20.-a Group theory)
%%%%%%%%%%%%%%%%%%%%%%%%%%%%%%%%%%%%%%%%%%%%%%%%%%%%%%%%%%%%%%%%%%%%5

%%------------------------------------------------------------------
\section{Introduction}
\label{sec:1}

\bigskip

Massive simulations in computer clusters have revealed that the control
parameter space of dissipative dynamical systems is riddled with stability
islands characterized
individually by periodic motions of ever increasing periods which accumulate,
exhibiting conspicuous and interesting regularities.
Even in systems governed by {\sl simple} polynomial maps, the number of periodic
orbits displays an explosive growth as a function of the
nonlinearity\cite{bg04}.
A few years ago it was realized that the nucleation of stability in classical
systems occurs in a variety of ways which normally involve the presence of
peak-doubling and peak-adding cascades extending over wide regions of the
control space. For instance, transitions among distinct stable oscillatory
phases may, or may not, be mediated by
parameter intervals, windows, of chaotic oscillations\cite{hg04,jg05,fgg7}.
%%%%%%%%%
Details of these and other novel regularities are summarized in recent surveys
concerning complexities observed in the accumulations of doubling and adding
cascades in laser systems see Ref.~\cite{ato}, in chemistry\cite{field},
in biochemical models\cite{fgg8,bio2}, and in the dynamics of
cancer\cite{cancer}.

Efficient algorithms for explicitly computing field isomorphisms and their
inverses for dynamical systems governed by polynomial maps are
central players for analytically assessing the aforementioned regularities
present in both theoretical and practical applications.
Polynomial maps are useful because metric properties of systems governed by
sets of nonlinear differential equations cannot yet be analytically
determined  due to the lack of methods to solve them exactly.
Since stability windows quickly become very narrow as oscillations increase,
accuracy is tantamount to exact analytical work.
One added advantage of polynomial maps is that the equations of motion
generated by them are always exact, not contaminated by the unavoidable
round-off and discretization
errors arising from numerical approximations of differential equations.

Currently, the common methods used for determining field isomorphism and
equivalence among equations of motion are:
i) a $p$-adic method reported by Zassenhaus and Liang and used\cite{zl69,cr74}
to study isomorphisms among quintic polynomials of all three
{\sl signatures}\cite{ha64b}
$(n,\ell)$, namely  $(1,2)$, $(3,1)$ and $(5,0)$, where $n$ refers to the number
of real roots while  $\ell$ refers to the number of pairs of complex roots;
ii) methods using factorization and linear algebra\cite{hp12};
iii) a systematic search of suitable Tschirnhausen transformations constrained
by some quantity of interest, usually polynomial discriminants.
This approach is efficient for systems with low-degree equations of
motion\cite{jg18}.

The purpose of this paper is to introduce an alternative method to detect
equivalence
among orbital equations of motion, namely Lagrange polynomial interpolation.
While standard methods look for isomorphisms focusing primarily on properties
of the number fields involved, the method based on polynomial interpolation
seeks isomorphisms
directly among irreducible polynomials, because they are the objects that arise
automatically as equations of motion governing periodic trajectories of
dynamical systems of algebraic origin\cite{sil2,kg,ks}.
After obtaining sets of equations of motion, the most important task in
dynamics is to
determined whether or not such equations are interconnected 
(i.e.~define or not the same number field), and when they are, to obtain
complete sets of explicit expressions for the transformations and
corresponding inverses interconnecting them.

%%------------------------------------------------------------------
\section{Polynomial interpolation as isomorphism detector}
\label{sec:2}

Although already used in 1779 by Edward Waring and an easy consequence
of a formula published in 1783 by Euler, ``Lagrange'' interpolation is a
result rediscovered by Lagrange in 1795 which is nowadays traditionally used in
numerical analysis for polynomial interpolation. The history of polynomial
interpolation is however considerably older than the works above, as reviewed
by Meijering\cite{me01}.

For a given set of $k$  pairs $(a_i, b_i)$, $i=1,\cdots,k$
with all $a_i$ distinct,
the interpolation polynomial is defined as the lowest degree polynomial that
assumes the value $b_i$ at the point $a_i$ for each $i$.
The interpolation polynomial is a linear combination
\[  L(x) = \sum_{i=0}^k  b_i \ell_i(x),   \]
where  
\[ \ell_i(x) = \frac{(x-a_1)\cdots(x-a_{i-1})(x-a_{i+1})\cdots(x-a_k)}
           {(a_i-a_1)\cdots(a_i-a_{i-1})(a_i-a_{i+1})\cdots(a_i-a_k)}
          = \prod_{j\neq i}\frac{x-a_j}{a_i-a_j}   \]
are the basis polynomials. %\cite{me01}.
The interpolator $L(x)$ has degree $k' \leq (k-1)$ and
$L(a_1)=b_1, \cdots, L(a_k)=b_k$.
In all cases of interest here, not only are the $a_i$ distinct but the $b_i$
are too, so we may reverse the interpolations.

Now, it is not difficult to see that by taking
$b_1=a_2, \cdots, b_{k-1}=a_k, b_k=a_1$ we obtain
$L(a_1)=a_2, \cdots, L(a_k)=a_1$ or, in other words, for this choice of $b_i$
the ``action'' of $L(x)$ is to induce a (cyclic) permutation among the
elements $a_1, \cdots, a_k$.
This is the basic observation that will be explored in the remainder of the
paper as an efficient detector of equivalence and isomorphism among
polynomial equations of motion.
In the applications considered here, the  elements $a_i$ and $b_i$ are usually
roots of algebraic orbital equations which have real coefficients and,
consequently, may be real or complex numbers.
Depending on the numerical values of $(a_i, b_i)$, the transformations
resulting from permutations of the $b_i$ will have coefficients defined by
real or complex numbers.
Isomorphisms of interest to us here are the ones characterized by rational
coefficients.
As seen in the examples in the next Section, in most cases, these interesting
isomorphisms involve just integer coefficients.

%===================================================================
\section{Applications}
\label{sec:3}

%---------------------------------------------------------------------
\subsection{Equivalences of Vandermonde's totally real cyclic quintic}

As a first application, we apply polynomial interpolation to find direct and
inverse transformations that establish the equivalence among a pair of cyclic
quintics of minimum discriminant $\Delta=14641=11^4$ originally considered
by Cohn\cite{cohn,jg18}, namely
\begin{eqnarray}
  V(x) &=& x^5-\phantom{2}x^4-4x^3+3x^2+3x-1,
             \qquad (\hbox{Vandermonde's quintic})
                                \label{VVV}\\
  G(x) &=& x^5+2x^4-5x^3-2x^2+4x-1.  \quad
                               \label{ggg}
\end{eqnarray}
The polynomial $V(x)$  represents a period-five orbital equation of motion for
at least three paradigmatic physical models in the so-called
{\sl generating partition limit}\cite{gl11}, namely
the quadratic map $x_{t+1} = 2-x_t^2$,
the H\'enon map $(x,y) \mapsto (2-x^2, y)$, and the canonical quartic
map\cite{jg94,jg95}, namely $x_{t+1} = (x_t^2-2)^2-2$.
For details see, e.g.~Refs.~\cite{res,haa}.
Apart from these fruitful applications, $V(x)$ is the celebrated quintic
solved by Vandermonde (1735-1796) using ``Lagrange'' resolvents a number of
years {\sl before}
Lagrange, and radical expressions at a time when it was still unknown
that a general solution of quintic equations was not
possible \cite{it,ni28,leb}.

%----------------------------------------------------------------------------
\begin{table}[!bht]
\centering
\tbl{The ten transformations interconnecting $V(x)$ and $G(x)$.}
{\begin{tabular}{@{}||ccccc | l||@{}} 
 \hline
  $b_1$ & $b_2$ & $b_3$ & $b_4$ & $b_5$ & Direct transforms: $V(x)\to G(x)$\\ %[0.5ex] 
 \hline\hline
   $g_4$ & $g_1$ & $g_2$ & $g_5$ & $g_3$ & $D_1 = -x^3+x^2+3x-2$\\  
 $g_5$ & $g_2$ & $g_4$ & $g_3$ & $g_1$ & $D_2 = -x^3+2x$\\
 $g_1$ & $g_5$ & $g_3$ & $g_2$ & $g_4$ & $D_3 = x^3-x^2-2x+1$\\
 $g_3$ & $g_4$ & $g_5$ & $g_1$ & $g_2$ & $D_4 = x^4-4x^2-x+2$\\
 $g_2$ & $g_3$ & $g_1$ & $g_4$ & $g_5$ & $D_5 = -x^4+x^3+4x^2-2x-3$\\
 \hline\hline
 $b_1$ & $b_2$ & $b_3$ & $b_4$ & $b_5$ & Inverse transforms: $G(x)\to V(x)$\\ %[0.5ex] 
 \hline\hline
  $v_2$ & $v_3$ & $v_5$ & $v_1$ & $v_4$ & $I_1 = 4x^4+10x^3-15x^2-15x+9$\\
  $v_2$ & $v_3$ & $v_1$ & $v_4$ & $v_5$ & $I_2 = x^4+2x^3-5x^2-3x+3$\\
   $v_4$ & $v_1$ & $v_2$ & $v_5$ & $v_3$ & $I_3 = -2x^4-5x^3+7x^2+7x-3$\\
   $v_3$ & $v_4$ & $v_5$ & $v_1$ & $v_2$ & $I_4 = -x^4-2x^3+5x^2+2x-3$\\  
   $v_3$ & $v_1$ & $v_2$ & $v_4$ & $v_5$ & $I_5 = -2x^4-5x^3+8x^2+9x-5$\\
 \hline
\end{tabular}}
\label{table:1}  
\end{table}
%----------------------------------------------------------------------------

To define basis polynomials and interpolations, we fix the roots of $V(x)$
and $G(x)$ in the following orders:
$$\begin{array}{lllll lllll}  
  v_1&\simeq -1.68,& \quad v_2&\simeq-0.83,& \quad v_3&\simeq0.28,& \quad
                           v_4&\simeq 1.30,& \quad v_5&\simeq 1.91,\\
  g_1&\simeq -3.22,& \quad g_2&\simeq -1.08,& \quad g_3&\simeq0.37,& \quad
                           g_4&\simeq0.54,& \quad g_5&\simeq1.39.
\end{array}
$$

Fixing $a_i = v_i$, we compute the five basis elements $\ell_i(x)$.
To check for the existence of transformations allowing the passage from $V(x)$
to $G(x)$ one needs to consider the 120 permutations of the roots $g_i$,
using each permuted set of roots as the points $b_i$.
This generates a set $\{ L_n(x)\}$ of transformations, for $n=1,\cdots, 120$.
Proceeding in this way, we find that, although the transformations may have
both real and complex coefficients, some permutations produce transformations
having just {\sl integer} or {\sl rational} coefficients.
For the passage from $V(x)$ to $G(x)$ we find five transformations given in
the upper part of Table \ref{table:1}, together with the root permutations
leading to them.
To obtain the inverse transformations, allowing the passage from $G(x)$ to
$V(x)$,
we fix $a_i=g_i$ and investigate the nature of the coefficients of the 120
transformations obtained by taking $b_i=v_i$ for all possible permutations
of the roots $v_i$.
As before, we find five inverse transformations, also listed in
Table  \ref{table:1}
with the root permutations leading to them.
To obtain the transformations we wrote a program for Maple 2018 (X86 64 LINUX)
running on a Dell XPS 13 notebook.
Recently, using systematic search of coefficients\cite{jg18}, it was possible to
discover the nine new transformations that had not been found with $p$-adic
methods.
Systematic coefficient search of the ten transformations required about 2.1
seconds and used 133.1 MB of memory.
In contrast, Lagrange interpolation has enabled these transformations to be
obtained in just 0.25 seconds and using only 13.7 MB of memory, a considerable
improvement.
This speedup of the classification of orbital points is, of course,
a desirable feature, that opens the possibility of investigating orbital
equations of considerably higher degrees.

While a systematic search of suitable Tschirnhausen transformations was able
to find the transformations in Table  \ref{table:1}, the interpolation
polynomials
introduced here have the advantage of revealing concomitantly, as a byproduct,
the nature of the action of the individual transformations on the roots.
We remark that some of the transformations in Table \ref{table:1} are
reducible, e.g.,
$D_2(x)$ is clearly reducible as also are $D_1$, $D_4$, $I_2$, and $I_3$.

%----------------------------------------------------------------------
\subsection{Hasse's problem: Equivalence of quintics with complex roots}

As a second application, we use polynomial interpolation to uncover four new
transformations, Eqs.~(\ref{new1})-(\ref{new4}),
providing a complete and ``symmetric solution'', i.e.~a solution providing both
{\sl direct} and {\sl inverse}  connections for a classical problem posed by
Hasse, who conjectured the possible existence of an isomorphism between three
quintics sharing a factor $47^2$ in their discriminant\cite{ha64,hl69}.
Such isomorphism was indeed confirmed by Zassenhaus and Liang\cite{zl69},
who used a $p$-adic method to uncover a pair of generating automorphisms of
the Hilbert class field over $\mathbb Q(\sqrt{-47})$.
Their pair of transformations is certainly enough to establish isomorphism
of the quintics  but, as mentioned, does not provide an unbiasedly balanced
and symmetric solution to Hasse's problem,
i.e.~a solution containing all possible {\sl direct} and {\sl inverse}
transformations among all polynomials involved.

Hasse's problem is concerned with relations between the zeros of three
quintic equations obtained by Weber \cite{web}, by Fricke \cite{fri}, and by
Hasse \cite{ha64,hl69}, while investigating class invariants
for modular equations with  discriminant $-47$.
The three quintics found by these authors are, respectively,
\begin{alignat*}{6}
  f_W  &=&\ &{x}^{5}-{x}^{3}-2\,{x}^{2}-2\,x-1,
  %  &\Delta_W &=&\ &47^2,
  &\theta_W &&\hbox{ the real root;}\\
%%  \label{eq2}\\
  f_F  &=&\ &{x}^{5}-{x}^{4}+{x}^{3}+{x}^{2}-2\,x+1,
%  &\Delta_F &=&\ &47^2.
  &\theta_F &&\hbox{ the real root;}\\
%%  \label{eq3}\\
  f_H  &=&\ &{x}^{5}+10\,{x}^{3}-235\,{x}^{2}+2610\,x-9353,\qquad
%  &\Delta_H &=&\ &47^2\cdot11^2\cdot5^{20},
  &\theta_H &&\hbox{ the real root.}
%%  \label{eq1} 
\end{alignat*}
As reported by Zassenhaus and Liang \cite{zl69}, Hasse asked whether or not
$\theta_W, \theta_F, \theta_H$ generate the same field.
And if so, how to express these roots in terms of each other?

Zassenhaus and Liang demonstrated that the polynomials indeed generate the same
field as manifest by the following transformations:
\begin{alignat*}{6}
  \theta_H &=&\ & 5\theta_W^2 -5\theta_W -2,\\ %\label{rel1}\\
  \theta_W &=&\ & -\theta_F^4 -2\theta_F +1. % \label{rel2}
\end{alignat*}
Proceeding as before, we find the additional root interconnections which read,
in the notation used by Zassenhaus and Liang:
\begin{eqnarray}
  \theta_F &=&\ -\theta_W^4 +\theta_W^3 + \theta_W +1,\label{new1}\\
  \theta_H &=&\   10\theta_F^4 - 5\theta_F^3 + 5\theta_F^2 + 10\theta_F - 12,\\
  \theta_W &=&\ \frac{1}{6875} \big(6\theta_H^4 + 23\theta_H^3 + 194\theta_H^2
                      - 1308\theta_H + 9821\big),\label{new3}\\
  \theta_F &=&\ -\frac{1}{6875} \big(\theta_H^4 + 13\theta_H^3
                        + 179\theta_H^2 + 717\theta_H - 444\big).\label{new4}
\end{eqnarray}
which may be easily verified.
Note the conspicuous presence of non-integer coefficients in Eqs.~(\ref{new3})
and (\ref{new4}), not a common occurrence in the literature.
Thus, polynomial interpolation is also able to deal with situations involving
not only real  but also complex roots.
Of course, not only the real roots but also all complex roots are properly
transformed by the same transformations above. When added to the known pair,
the four new transformations reported here solve Hasse's problem completely
and symmetrically, with no bias.

%----------------------------------------------------------------------------
\begin{table}[!bht]
\centering
\tbl{Direct transformations among the $s_i(x)$ of
  Eqs.~(\ref{zla})-(\ref{zld}).}
{\begin{tabular}{@{}||cccccc|l||@{}} 
\hline\hline
 $b_1$ & $b_2$ & $b_3$ & $b_4$ & $b_5$ & $b_6$ &Transforms from $s_1(x) \to s_2(x)$\\ [0.5ex]
\hline\hline
1& 4& 2& 6& 5& 3& $2x^5-x^4-14x^3-4x^2+10x+2$\\
2& 1& 3& 4& 6& 5& $-3x^5+x^4+21x^3+9x^2-11x-4$\\
3& 2& 5& 1& 4& 6& $-6x^5+2x^4+43x^3+17x^2-28x-7$\\
4& 6& 1& 5& 3& 2& $4x^5-2x^4-28x^3-7x^2+18x+2$\\
5& 3& 6& 2& 1& 4& $3x^5-22x^3-15x^2+12x+6$\\
6& 5& 4& 3& 2& 1& $-x$\\
\hline \hline
 $b_1$ & $b_2$ & $b_3$ & $b_4$ & $b_5$ & $b_6$ & Transforms from $s_1(x) \to s_3(x)$ \\ [0.5ex] 
\hline\hline
1& 2& 5& 3& 6& 4& $-7x^5+2x^4+50x^3+22x^2-30x-8$\\
2& 3& 1& 6& 4& 5& $6x^5-2x^4-43x^3-17x^2+29x+7$\\
3& 6& 2& 4& 5& 1& $-x^4+x^3+6x^2-x-2$\\
4& 5& 6& 1& 2& 3& $-x^5+x^4+7x^3-2x^2-7x+2$\\
5& 1& 4& 2& 3& 6& $-2x^5+x^4+14x^3+4x^2-9x-2$\\
6& 4& 3& 5& 1& 2& $4x^5-x^4-29x^3-13x^2+18x+5$\\
\hline\hline
 $b_1$ & $b_2$ & $b_3$ & $b_4$ & $b_5$ & $b_6$ &Transforms from $s_1(x) \to s_4(x)$\\ [0.5ex]
\hline\hline
1& 3& 4& 2& 6& 5& $-4x^5+x^4+29x^3+13x^2-18x-5$\\
2& 6& 3& 5& 4& 1& $2x^5-x^4-14x^3-4x^2+9x+2$\\
3& 2& 1& 6& 5& 4& $x^5-x^4-7x^3+2x^2+7x-2$\\
4& 1& 5& 3& 2& 6& $x^4-x^3-6x^2+x+2$\\
5& 4& 6& 1& 3& 2& $-6x^5+2x^4+43x^3+17x^2-29x-7$\\
6& 5& 2& 4& 1& 3& $7x^5-2x^4-50x^3-22x^2+30x+8$\\
\hline\hline
 $b_1$ & $b_2$ & $b_3$ & $b_4$ & $b_5$ & $b_6$ &Transforms from $s_2(x) \to s_3(x)$\\ [0.5ex]
\hline\hline
1& 5& 4& 2& 6& 3& $-x^4-x^3+6x^2+x-2$\\
2& 1& 5& 3& 4& 6& $-4x^5-x^4+29x^3-13x^2-18x+5$\\
3& 2& 1& 6& 5& 4& $x^5+x^4-7x^3-2x^2+7x+2$\\
4& 6& 3& 5& 2& 1& $7x^5+2x^4-50x^3+22x^2+30x-8$\\
5& 4& 6& 1& 3& 2& $-6x^5-2x^4+43x^3-17x^2-29x+7$\\
6& 3& 2& 4& 1& 5& $2x^5+x^4-14x^3+4x^2+9x-2$\\
\hline\hline
$b_1$ & $b_2$ & $b_3$ & $b_4$ & $b_5$ & $b_6$ &Transforms from $s_2(x) \to s_4(x)$\\ [0.5ex]
\hline\hline
1& 4& 5& 3& 6& 2& $-2x^5-x^4+14x^3-4x^2-9x+2$\\
2& 3& 1& 6& 4& 5& $6x^5+2x^4-43x^3+17x^2+29x-7$\\
3& 1& 4& 2& 5& 6& $-7x^5-2x^4+50x^3-22x^2-30x+8$\\
4& 5& 6& 1& 2& 3& $-x^5-x^4+7x^3+2x^2-7x-2$\\
5& 6& 2& 4& 3& 1& $4x^5+x^4-29x^3+13x^2+18x-5$\\
6& 2& 3& 5& 1& 4& $x^4+x^3-6x^2-x+2$\\
\hline\hline
$b_1$ & $b_2$ & $b_3$ & $b_4$ & $b_5$ & $b_6$ &Transforms from $s_3(x) \to s_4(x)$\\ [0.5ex]
\hline\hline
1& 3& 2& 5& 4& 6& $2x^5-6x^4-7x^3+8x^2+3x-1$\\
2& 6& 5& 1& 3& 4& $-2x^4+6x^3+7x^2-8x-2$\\
3& 2& 6& 4& 1& 5& $x^4-2x^3-6x^2+2$\\
4& 1& 3& 6& 5& 2& $-2x^5+7x^4+4x^3-12x^2+3x+2$\\
5& 4& 1& 2& 6& 3& $-x^3+3x^2+3x-3$\\
6& 5& 4& 3& 2& 1& $-x$\\
\hline\hline
\end{tabular}}
\label{table:2}  
\end{table}
%----------------------------------------------------------------------------

%----------------------------------------------------------------------------
\begin{table}[!bht]
\centering
\tbl{Inverse transformations among the $s_i(x)$ of
  Eqs.~(\ref{zla})-(\ref{zld}). \ \ }
{\begin{tabular}{@{}||cccccc|l||@{}} 
\hline\hline
 $b_1$ & $b_2$ & $b_3$ & $b_4$ & $b_5$ & $b_6$ &Transforms from $s_2(x) \to s_1(x)$\\ [0.5ex]
\hline\hline
1& 3& 6& 2& 5& 4& $-6x^5-2x^4+43x^3-17x^2-2x+7$\qquad\\
2& 1& 3& 4& 6& 5& $-3x^5-x^4+21x^3-9x^2-11x+4$\\
3& 6& 5& 1& 4& 2& $3x^5-22x^3+15x^2+12x-6$\\
4& 2& 1& 5& 3& 6& $2x^5+x^4-14x^3+4x^2+10x-2$\\
5& 4& 2& 6& 1& 3& $4x^5+2x^4-28x^3+7x^2+18x-2$\\
6& 5& 4& 3& 2& 1& $-x$\\
\hline\hline
 $b_1$ & $b_2$ & $b_3$ & $b_4$ & $b_5$ & $b_6$ &Transforms from $s_3(x) \to s_1(x)$\\ [0.5ex]
\hline\hline
1& 2& 4& 6& 3& 5& $x^5-2x^4-6x^3-x^2+4x+2$\\
2& 4& 5& 3& 1& 6& $x^5-3x^4-3x^3+3x^2-x$\\
3& 1& 2& 5& 6& 4& $-x^5+3x^4+3x^3-3x^2+2x$\\
4& 5& 6& 1& 2& 3& $-x^5+2x^4+7x^3-2x^2-7x+1$\\
5& 6& 3& 2& 4& 1& $x^5-4x^4-x^3+9x^2-2x-2$\\
6& 3& 1& 4& 5& 2& $-x^5+4x^4-6x^2+4x$\\
\hline\hline
 $b_1$ & $b_2$ & $b_3$ & $b_4$ & $b_5$ & $b_6$ &Transforms from $s_4(x) \to s_1(x)$\\[0.5ex]
\hline\hline
1& 4& 2& 3& 6& 5& $-x^5-4x^4+x^3+9x^2+2x-2$\\
2& 5& 4& 1& 3& 6& $x^5+4x^4-6x^2-4x$\\
3& 2& 1& 6& 5& 4& $x^5+2x^4-7x^3-2x^2+7x+1$\\
4& 6& 5& 2& 1& 3& $x^5+3x^4-3x^3-3x^2-2x$\\
5& 3& 6& 4& 2& 1& $-x^5-2x^4+6x^3-x^2-4x+2$\\
6& 1& 3& 5& 4& 2& $-x^5-3x^4+3x^3+3x^2+x$\\
\hline\hline
 $b_1$ & $b_2$ & $b_3$ & $b_4$ & $b_5$ & $b_6$ &Transforms from $s_3(x) \to s_2(x)$\\[0.5ex]
\hline\hline
1& 4& 6& 3& 2& 5& $x^5-4x^4+6x^2-4x$\\
2& 1& 4& 5& 3& 6& $-x^5+4x^4+x^3-9x^2+2x+2$\\
3& 2& 1& 6& 5& 4& $x^5-2x^4-7x^3+2x^2+7x-1$\\
4& 6& 5& 2& 1& 3& $x^5-3x^4-3x^3+3x^2-2x$\\
5& 3& 2& 4& 6& 1& $-x^5+3x^4+3x^3-3x^2+x$\\
6& 5& 3& 1& 4& 2& $-x^5+2x^4+6x^3+x^2-4x-2$\\
\hline\hline
 $b_1$ & $b_2$ & $b_3$ & $b_4$ & $b_5$ & $b_6$ &Transforms from $s_4(x) \to s_2(x)$\\[0.5ex]
\hline\hline 
1& 6& 4& 2& 3& 5& $x^5+3x^4-3x^3-3x^2-x$\\
2& 4& 1& 3& 5& 6& $x^5+2x^4-6x^3+x^2+4x-2$\\
3& 1& 2& 5& 6& 4& $-x^5-3x^4+3x^3+3x^2+2x$\\
4& 5& 6& 1& 2& 3& $-x^5-2x^4+7x^3+2x^2-7x-1$\\
5& 2& 3& 6& 4& 1& $-x^5-4x^4+6x^2+4x$\\
6& 3& 5& 4& 1& 2& $x^5+4x^4-x^3-9x^2-2x+2$\\
\hline\hline
 $b_1$ & $b_2$ & $b_3$ & $b_4$ & $b_5$ & $b_6$ &Transforms from $s_4(x) \to s_3(x)$\\[0.5ex]
\hline\hline
1& 3& 2& 5& 4& 6& $2x^5+6x^4-7x^3-8x^2+3x+1$\\
2& 6& 3& 1& 5& 4& $-x^4-2x^3+6x^2-2$\\
3& 4& 6& 2& 1& 5& $2x^4+6x^3-7x^2-8x+2$\\
4& 1& 5& 6& 3& 2& $-x^3-3x^2+3x+3$\\
5& 2& 1& 4& 6& 3& $-2x^5-7x^4+4x^3+12x^2+3x-2$\\
6& 5& 4& 3& 2& 1& $-x$\\
\hline\hline
\end{tabular}}
\label{table:3}  
\end{table}
%----------------------------------------------------------------------------

As described by Pohst\cite{po94}, Hasse's problem played an important role
in establishing computational algebraic number theory at a time when
computations of all kinds were taboo. In the early 1960s, Zassenhaus developed
algorithmic means for number theoretical experiments in algebraic number
theory. A major success was the proof of the isomorphism of the three quintic
fields which occurred as candidates for the real subfield of the Hilbert class
field of $\mathbb Q(\sqrt{-47})$. This problem, pointed out by Hasse, could
not be solved by theoretical methods. It was the numerical solution
by Zassenhaus and Liang\cite{zl69} that gave major credit to methods in
constructive algebraic number theory.

%\vspace{-1.5truecm}

%--------------------------------------------------------------------------
\subsection{Equivalence among totally real sextics with small coefficients}

A very interesting and much studied class of equations of motion involves
sextic polynomials \cite{eg06}.
Their splitting fields may contain quadratic and cubics subfields or no
subfields at all.
Thus, sextics may appear generically as orbital clusters   
algebraically entangling orbits into distinct groups of periodicity.
Of particular interest  is knowledge concerning cyclic sextics.
The minimum discriminant of sextics, $300,125=5^3\cdot7^4$, was found by
Liang and Zassenhaus for the polynomial $s_1(x)$ a totally real
cyclic sextic \cite{lz77}:
\begin{eqnarray}
      s_1(x) &=& x^6-x^5-7x^4+2x^3+7x^2-2x-1,\label{zla}\\
      s_2(x) &=& x^6+x^5-7x^4-2x^3+7x^2+2x-1,\label{zlb}\\
      s_3(x) &=& x^6-2x^5-7x^4+2x^3+7x^2-x-1,\label{zlc}\\
      s_4(x) &=& x^6+2x^5-7x^4-2x^3+7x^2+x-1.\label{zld}
\end{eqnarray}      
The same minimum discriminant is also shared by $s_2(x), s_3(x), s_4(x)$,
simple reincarnations of $s_1(x)$ after the substitutions $x\to\pm x$
and $x\to \pm1/x$ and suitable simplifications.
%%%
Curiously, the roots of the above polynomials are interconnected in subtle ways
by a multitude of transformations, given in Tables \ref{table:2} and
\ref{table:3}.
The values of the $b_i$ given in these tables indicate the {\sl order} of the
roots needed to find the corresponding transformations.
As before for Vandermonde's quintic, we assume the roots of the $s_i(x)$ to
be ordered from the smallest to the largest.
Thus, denoting by $s_1^{(i)}$, $i=1,\dots,6$,  the ordered roots of $s_1(x)$,
the transformation shown in the first line of Table \ref{table:2} is obtained
when fixing $b_i = s_1^{(i)}$, and so on.

Similarly as for the $s_i(x)$, polynomials characterized by totally real cyclic
sextics and {\sl second lowest}  discriminant, namely $371,293$, are the
following:
\begin{eqnarray*}
      t_1(x) &=& {x}^{6}-{x}^{5}-5\,{x}^{4}+4\,{x}^{3}+6\,{x}^{2}-3\,x-1,\\
      t_2(x) &=& {x}^{6}+{x}^{5}-5\,{x}^{4}-4\,{x}^{3}+6\,{x}^{2}+3\,x-1,\\
      t_3(x) &=& {x}^{6}-3\,{x}^{5}-6\,{x}^{4}+4\,{x}^{3}+5\,{x}^{2}-x-1,\\
      t_4(x) &=& {x}^{6}+3\,{x}^{5}-6\,{x}^{4}-4\,{x}^{3}+5\,{x}^{2}+x-1,\\
      t_5(x) &=& {x}^{6}-2\,{x}^{5}-7\,{x}^{4}+6\,{x}^{3}+5\,{x}^{2}-5\,x+1,\\
      t_6(x) &=& {x}^{6}+2\,{x}^{5}-7\,{x}^{4}-6\,{x}^{3}+5\,{x}^{2}+5\,x+1,\\
      t_7(x) &=& {x}^{6}-5\,{x}^{5}+5\,{x}^{4}+6\,{x}^{3}-7\,{x}^{2}-2\,x+1,\\
      t_8(x) &=& {x}^{6}+5\,{x}^{5}+5\,{x}^{4}-6\,{x}^{3}-7\,{x}^{2}+2\,x+1.
\end{eqnarray*}      
As for the $s_i(s)$, note the conspicuous presence of two groups of four
elements arising from the substitutions $x\to\pm x$ and $x\to \pm1/x$
and suitable simplifications.

%----------------------------------------------------------------------------
\begin{table}[!bht]
\centering
\tbl{Bridges among sextics arising in orbital clusters of the H\'enon
  Hamiltonian repeller. Note the non-integer coefficients in
  the first column, not a common occurrence in the literature.}
{\begin{tabular}{@{}||l|l||@{}} 
\hline\hline
 Direct  & Inverse \\ [0.5ex]
\hline\hline
$X\to Y:$ \   $-\tfrac{1}{2}x^2+3$
                & $Y\to X:$ \  $-x^4+2x^3+9x^2-13x-16$\\[0.7ex]
$X\to Z:$ \  $-\tfrac{1}{4}x^4+3x^2-x-3$
                & $Z\to X:$ \ $-x^4 -2x^3 +11x^2 +11x -30$\\[0.7ex]
$Y\to Z:$ \  $x^4-2x^3-10x^2+13x+22$ 
                & $Z\to Y:$ \ $-x^2-x+6=-(x+3)(x-2)$\\[0.7ex]
\hline
$U\to W:$ \  $-\tfrac{1}{4}x^4 +3x^2-x-3$
& $W\to U:$ \ $-4x^5 - 6x^4 +63x^3 +74x^2 -234x -219$\\[0.7ex]
\phantom{$U\to W:$ \ }$-\tfrac{1}{2}x^2 +3$
& \phantom{$W\to U:$  }
     $\phantom{-}2x^5+3x^4-32x^3-38x^2+121x+116$\\[0.7ex]
\hline\hline
\end{tabular}}
\label{table:amalga}  
\end{table}
%----------------------------------------------------------------------------

To each pair of polynomials $t_i(x)$ corresponds a set of six transformations
analogous to the ones in Tables \ref{table:2} and \ref{table:3}, resulting
from similar root permutations.
The total number of such transformations is 64, too many to be recorded
here explicitly.
However, the polynomials $t_i(x)$ allow them to be obtained easily if so
desired, along with the proper root permutations leading to them.

%---------------------------------------------------------------------------
\subsection{Equivalent totally real cyclic sextics with larger coefficients}

Table \ref{table:amalga} reports isomorphisms having a direct bearing on the
inner workings of the H\'enon Hamiltonian repeller\cite{eg06}.
As seen in the previous Section, sextics leading to {\sl minimal discriminants}
tend to have comparatively small coefficients.
However, in real-life applications the coefficients are not always so small,
for instance the cluster $\mathbb A^{(5)}(x) = X(x) Y^2(x) Z^2(x)$
defining the orbital coordinates of six period-five trajectories of the H\'enon
Hamiltonian repeller involves totally real sextics with larger coefficients:
%  and Galois group 6T16:
%\begin{eqnarray*}
\begin{alignat*}{6}
 X(x) &=& \  &{x}^{6}-2\,{x}^{5}-14\,{x}^{4}+24\,{x}^{3}
   +32\,{x}^{2}-16\,x-8,  &\quad\Delta_X&=2^{18}\cdot 31\cdot 241 \cdot 389,\\
 Y(x) &=&  &{x}^{6}-2\,{x}^{5}-16\,{x}^{4}+26\,{x}^{3}
    +81\,{x}^{2}-84\,x- 125, &\Delta_Y&=2^{6}\cdot 31\cdot 241 \cdot 389,\\
 Z(x) &=&  &{x}^{6}+2\,{x}^{5}-16\,{x}^{4}-22\,{x}^{3}
 +85 \,{x}^{2}+60\,x-151, &\Delta_Z&=2^{6}\cdot 31\cdot 241 \cdot 389.
\end{alignat*}
Similarly, a trio of period-five orbits is amalgamated into
$\mathbb B^{(5)}(x) = U(x)W^2(x)$,
defining the eighteen orbital coordinates as roots of a pair of totally real
sextics\cite{eg06}, with Galois group 6T7:
%\begin{eqnarray*}
\begin{alignat*}{6}  
  U(x) &=& \ &x^6       -22x^4 +8x^3  +124x^2 -88x  -32,
        &\quad&\Delta_U&=\ &2^{18}\cdot 3^4\cdot 659^2,  \\
  W(x) &=& \ &x^6 +4x^5 -12x^4 -58x^3 +12x^2  +202x +139,
        &\quad&\Delta_W&=\ &2^{6}\cdot 3^4\cdot 659^2. 
\end{alignat*}
%\end{eqnarray*}

%----------------------------------------------------------------------------
\begin{table}[!bht]
\centering
\tbl{Some representative {\sl bridges} that allow direct and inverse passage
  among polynomials of the families $f_i(x)$ and $g_i(x)$.
  There are no connections between the $f_i(x)$ and $g_i(x)$ and vice-versa,
  despite the fact that
  they all share the same discriminant $810,448=2^2\cdot 37^3$.}
{\begin{tabular}{@{}||l|l||@{}} 
\hline\hline
 Direct  & Inverse \\ [0.5ex]   %  $\\[0.7ex]
\hline\hline
$f_1\to f_2:$ \   $-x$  
& $f_2\to f_1:$ \ $-x$ \\[0.7ex]
\phantom{$f_1\to f_2:$} \   $x - 1$
& \phantom{$f_2\to f_1:$} \ $x+1$\\[0.7ex]
\phantom{$f_1\to f_2:$} \   $-x^5 + 2x^4 + 4x^3 - 6x^2 - 4x + 2$
& \phantom{$f_2\to f_1:$} \ $-x^5 - 3x^4 + 2x^3 + 8x^2 - x - 2$\\[0.7ex]
\phantom{$f_1\to f_2:$} \   $-x^5 + 3x^4 + 2x^3 - 8x^2 - x + 2$
& \phantom{$f_2\to f_1:$} \ $-x^5 - 2x^4 + 4x^3 + 6x^2 - 4x - 2$\\[0.7ex]
\phantom{$f_1\to f_2:$} \   $x^5 - 3x^4 - 2x^3 + 8x^2 + x - 3$
& \phantom{$f_2\to f_1:$} \ $x^5 + 2x^4 - 4x^3 - 6x^2 + 4x + 3$\\[0.7ex]
\phantom{$f_1\to f_2:$} \   $x^5 - 2x^4 - 4x^3 + 6x^2 + 4x - 3$
& \phantom{$f_2\to f_1:$} \ $x^5 + 3x^4 - 2x^3 - 8x^2 + x + 3$\\[0.7ex]
\hline\hline  %------------------------
$g_1\to g_2:$ \   $-x$
& $g_2\to g_1:$ \ $-x$\\[0.7ex]
\phantom{$g_1\to g_2:$} \   $x^5 - 5x^4 + 8x^3 - 9x^2 + 8x - 5$
& \phantom{$g_2\to g_1:$} \ $x^5 + 5x^4 + 8x^3 + 9x^2 + 8x + 5$\\[0.7ex]
\hline  %------------------------
$g_1\to g_3:$ \   $-x^5 + 4x^4 - 4x^3 + 5x^2 - 3x + 2$
& $g_3\to g_1:$ \ $-x^5 + 3x^4 - 6x^3 + 7x^2 - 2x$\\[0.7ex]
\phantom{$g_1\to g_3:$} \   $x^5 - 4x^4 + 4x^3 - 5x^2 + 3x - 1$
& \phantom{$g_3\to g_1:$} \ $x^5 - 2x^4 + 4x^3 - 3x^2 - x + 1$\\[0.7ex]
\hline  %------------------------
$g_1\to g_4:$ \   $-x^5 + 4x^4 - 4x^3 + 5x^2 - 3x + 1$
& $g_4\to g_1:$ \ $-x^5 - 2x^4 - 4x^3 - 3x^2 + x + 1$\\[0.7ex]
\phantom{$g_1\to g_4:$} \   $x^5 - 4x^4 + 4x^3 - 5x^2 + 3x - 2$
& \phantom{$g_4\to g_1:$} \ $x^5 + 3x^4 + 6x^3 + 7x^2 + 2x$\\[0.7ex]
\hline  %------------------------
$g_1\to g_5:$ \   $-x$
& $g_5\to g_1:$ \ $-x$\\[0.7ex]
\phantom{$g_1\to g_5:$} \   $x^5 - 5x^4 + 8x^3 - 9x^2 + 8x - 4$
& \phantom{$g_5\to g_1:$} \ $x^5 - 2x^3 + 5x^2 - x + 2$\\[0.7ex]
\hline  %------------------------
$g_1\to g_6:$ \   $x + 1$
& $g_6\to g_1:$ \ $x - 1$\\[0.7ex]
\phantom{$g_1\to g_6:$} \   $-x^5 + 5x^4 - 8x^3 + 9x^2 - 8x + 4$
& \phantom{$g_6\to g_1:$} \ $-x^5 + 2x^3 + 5x^2 + x + 2$\\[0.7ex]
%%%%
\hline\hline
\end{tabular}}
\label{table:duplo}  
\end{table}
%----------------------------------------------------------------------------

Table \ref{table:amalga} shows that the orbits algebraically entangled
together to form
the above pair of orbital clusters have their coordinates related by
{\sl simple transformations} that, surprisingly,
{\sl allow back and forth passage among seemingly distinct orbits.}
Unfortunately, the Galois group of $X(x)$, $Y(x)$, $Z(x)$, $U(x)$, and $W(x)$
is the symmetric group, meaning that these sextics cannot be solved in terms
of a finite number of radical extractions and elementary arithmetic operations.
But the transformations interconnecting these sextics show clearly that
knowledge of just two sets of six roots is enough to interconnect in
phase-space all orbital points of the
equations algebraically entangled in each cluster.
The sextic trio  of cluster  $\mathbb A^{(5)}(x)$ is not isomorphic
to the pair of sextics of cluster $\mathbb B^{(5)}(x)$.
Table \ref{table:amalga} contains transformations with non-integer coefficients,
something that we have not been able to find in the literature.

%--------------------------------------------------------------------------
\subsection{Equivalence among distinct families of isodiscriminant sextics}

As a quite remarkable final example,
we consider a family of ten totally real sextics sharing the same discriminant,
$810,448=2^2\cdot 37^3$, but
formed by two {\sl non-isomorphic} families $f_i(x)$ and $g_i(x)$,
defined in Eqs.~(\ref{eq1})-(\ref{eq10}).
The four sextics $f_i(x)$ are isomorphic among themselves, as also are
the six $g_i(x)$.
However, none of the $f_i(x)$  is isomorphic to any of the $g_i(x)$
and vice-versa, despite the fact that they all share the same discriminant.
In the notation of Butler and McKay\cite{bm83} adopted by Maple, the Galois
group of the $f_i(x)$ is 6T1, a cyclic semiabelian group,
while the group of the $g_i(x)$ is 6T8, a solvable, semiabelian group.

\begin{eqnarray}
f_1(x) &=&  x^6 -3x^5-2x^4+9x^3-5x+1,\label{eq1}\\
f_2(x) &=&  x^6 +3x^5-2x^4-9x^3+5x+1,\\
f_3(x) &=&  x^6 -5x^5+9x^3-2x^2-3x+1,\\
f_4(x) &=&  x^6 +5x^5-9x^3-2x^2+3x+1,
\end{eqnarray}
\begin{eqnarray}
g_1(x) &=& x^6 -5x^5+8x^4-9x^3+8x^2-5x+1,\\
g_2(x) &=& x^6 +5x^5+8x^4+9x^3+8x^2+5x+1,\\
g_3(x) &=& {x}^{6} -3\,{x}^{5}+6\,{x}^{4}-7\,{x}^{3}+2\,{x}^{2}+x-1,\\
g_4(x) &=& x^6 +3x^5+6x^4+7x^3+2x^2-x-1,\\
g_5(x) &=& {x}^{6} -{x}^{5}-2\,{x}^{4}+7\,{x}^{3}-6\,{x}^{2}+3\,x-1,\\
g_6(x) &=& {x}^{6} +{x}^{5}-2\,{x}^{4}-7\,{x}^{3}-6\,{x}^{2}-3\,x-1.\label{eq10}
\end{eqnarray}
Once again, note in Eqs.~(\ref{eq1})-(\ref{eq10})
two groups of four elements underlying the substitutions
$x\to\pm x$ and $x\to \pm1/x$, and the presence of two {\sl outliers},
namely the reciprocal polynomials $g_1(x)$ and $g_2(x)$.

Proceeding as before,
from the roots of Eqs.~(\ref{eq1})-(\ref{eq10}) one may easily obtain the large
set of transformations allowing back and forth passage, local to the global,
among both groups of sextics.
There is a total of six transformations connecting each pair of $f_i(x)$ but
just two transformations connecting pairs of $g_i(x)$.
The complete set of transformations is omitted here, with just a few
representative ones being given in Table \ref{table:duplo}.

%===================================================================
\section{Conclusions and outlook}
\label{sec:fim}

This paper has shown that Lagrange interpolation works as an efficient detector
of equivalence and isomorphism among orbital equations of motion of algebraic
dynamical systems governed by discrete-time mappings. 
This is a startling new application for a well-known interpolation technique of
numerical analysis. Here, it is not used to {\sl approximate} anything but,
instead,
as means of obtaining {\sl exact analytical expressions for isomorphisms}.
We found polynomial interpolation to efficiently detect equivalences among
equations
of any signature, i.e.~among polynomials having only real roots or not.
The method is simple to implement and very fast.
We anticipate polynomial interpolation to be a helpful tool to locate
equivalences among the huge number\cite{bg04} of orbital equations in systems
of algebraic origin and polynomials in general.
In particular, it should help to uncover equivalences among, e.g., the
complicated
{\sl amalgamation} polynomial clusters arising in the Hamiltonian repeller
limit of the H\'enon map\cite{eg06}, and among orbits of the Pincherle map,
a paradigmatic map underlying the operating kernel of the so-called
{\sl chaotic computer}\cite{res,cc2,cc3,cc4}.

%\medskip%\medskip
%%------------------------------------------------------------------
\section*{Acknowledgments}
JACG thanks helpful email exchanges with J.~Voight and J.C.~Interlando.
The latter brought Ref.~\cite{hp12} to our attention.
OJB acknowledges the partial support of the Funda\c c\~ao para a Ci\^encia
e Tecnologia,
project UID/MAT/04721/2013 (Centro de An\'alise Funcional, Estruturas
Lineares e Aplica\c c\~oes - Grupo de Estruturas Lineares, Alg\'ebricas e
Combinat\'orias, Universidade de Lisboa, Portugal).
JACG was supported by CNPq, Brazil.
This work was also supported by the
Max-Planck Institute for the Physics of Complex Systems, Dresden,
in the framework of the Advanced Study Group on Optical Rare Events.

%\appendix

%\section{Appendices}
%Appendices should be used only when absolutely necessary. They
%should come before the References. If there is more than one
%appendix, number them alphabetically. Number displayed equations
%occurring in the Appendix in this way, e.g.~(\ref{app1}),\break

%\section*{References}
%References are to be listed in the order cited in the text in Arabic
%numerals.  They can be typed in superscripts after punctuation marks,
%e.g.,~``$\ldots$ in the statement.\cite{1}'' or used directly,
%e.g.,~``see Ref.~\refcite{1} for examples''. Please list using the
%style shown in the following examples.  For journal names, use the
%standard abbreviations or spell in full. Typeset the references in 9 pt
%roman with baselineskip of 11 pt.

%\begin{thebibliography}{000} %for 3 digits
%\begin{thebibliography}{00}  %for 2 digits
%\begin{thebibliography}{0}   %for 1 digit

\end{document}